\documentclass[12pt]{iopart}
\usepackage{iopams}  

\begin{document}

\newcommand\mythesection{\arabic{section}}

\newcommand\mybeginsubeqs[1]{
   \refstepcounter{equation}%
   \if ?#1?\else\label{#1}\fi%
   \setcounter{eqnval}{\value{equation}}%
   \setcounter{equation}{0}%
   \def\theequation{\mythesection.\arabic{eqnval}{\it\alph{equation}}}%
   }

\def\myendsubeqs{
   \def\theequation{\mythesection.\arabic{equation}}%
   \setcounter{equation}{\value{eqnval}}%
   }

\def\theequation{\mythesection.\arabic{equation}}
\newtheorem{proposition}{Proposition}[section]
\newtheorem{propositionA}{Proposition}[subsection]

			
\renewcommand\vec[1]{\boldsymbol{#1}}
\def\tfrac#1#2{{\textstyle{#1\over#2}}}

\newcommand{\myu}{\mathop{u}} 
\newcommand{\myw}{\mathop{w}} 
\newcommand{\myv}{\mathop{\hat{u}}} 
\newcommand{\myin}[1]{\vec{n} \in \Lambda^{#1}} 
\newcommand{\myijk}{\{i,j,k\} = \{1,2,3\}} 

\newcommand{\mylambda}[2][]{\if ?#1? \lambda_{#2} \else \lambda_{#2}^{#1}\fi} 
\newcommand{\mymatrix}[1]{\mathsf{#1}} 
\newcommand{\myconst}[2]{\mathop{#1_{#2}}}
\newcommand{\myalhconst}[2]{\mathop{#1(#2)}}

\newcommand\myShift[2][T]{\mathop{\mathbb{#1}}\nolimits_{#2}}
\newcommand\myShifted[3][0]{\ifcase #1
   \mathbb{T}_{#2}{#3} \or                     
   \left(\mathbb{T}_{#2}{#3}\right) \or        
   \mathbb{T}_{#2}^{-1}{#3} \or                
   \left(\mathbb{T}_{#2}^{-1}{#3}\right) \else 
   #3[#2] 
 \fi}


\title{Solitons of a simple nonlinear model on the cubic lattice.}
\author{V.E. Vekslerchik}
\address{
  Usikov Institute for Radiophysics and Electronics \\
  12, Proskura st., Kharkov, 61085, Ukraine 
}
\ead{vekslerchik@yahoo.com}
\submitto{\JPA}


\begin{abstract}
We study a simple nonlinear model defined on the cubic lattice. 
We propose a bilinearization scheme for the field equations and 
demonstrate that the resulting system is closely related to the well-studied 
integrable models, such as the Hirota bilinear difference equation and the 
Ablowitz-Ladik system. This result is used to derive the two sets of the 
N-soliton solutions. 
\end{abstract}

\section{Introduction.}

In this paper we try to extend the area of application of the direct methods 
of the soliton theory. 
We show that the approaches developed in our previous works 
\cite{V16a,V16b} can be used to obtain a wide range of explicit solutions 
for nonlinear lattice models in three dimensions.

The model which we study seems to be new. We do not address the question of 
its integrability. Instead, we show that by means of elementary 
transformations one can reduce it to the well-studied integrable 
systems (the Ablowitz-Ladik hierarchy (ALH) \cite{AL75} and the Hirota 
bilinear difference equation (HBDE) \cite{H81}). After that, we can use the 
standard techniques to derive solutions (or even use those  already known) 
for our model, which are difficult to obtain by means of the straightforward 
approaches. 

In this paper we restrict ourselves to the N-soliton solutions which are not 
only interesting in themselves, but also hint at (but surely do not prove) 
the integrability of the model.

\section{Model.}

The model which we study in this paper describes the scalar fields defined at 
the vertices of the cubic lattice with the logarithmic interaction between the 
nearest neighbours,
\begin{equation}
  \mathcal{S} 
  = 
  \sum\limits_{\vec{n}' \sim \vec{n}''} 
    \Gamma_{\vec{n}'\vec{n}''} 
    \ln\left[ 
      1 + 
      \myu(\vec{n}') \myu(\vec{n}'')  
    \right] 
\label{def:lnn}
\end{equation}
where $\vec{n}' \sim \vec{n}''$ means that vectors $\vec{n}'$ and $\vec{n}''$ 
point to adjacent nodes of the lattice and $\Gamma_{\vec{n}'\vec{n}''}$ are 
interaction constants that depend on the type (orientation) of the edge 
(see below).
In more detail, we present the cubic lattice as 
\begin{equation}
  \Lambda 
  = 
  \left\{ 
    \vec{n} = \sum\limits_{i=1}^{3} n_{i} \vec{e}_{i}, 
    \quad
    n_{i} \in \mathbb{Z}, \; 
    \vec{e}_{i} \in \mathbb{R}^{3}
  \right\} 
\label{def:lattice}
\end{equation}
and define the model by 
\begin{equation}
  \mathcal{S} 
  = 
  \sum\limits_{\vec{n} \in \Lambda} \, 
  \sum\limits_{i=1}^{3} 
    \Gamma_{i} 
    \ln\left[ 
      1 + 
      \myu(\vec{n}) \myu(\vec{n}+\vec{e}_{i})  
    \right] 
\label{def:model}
\end{equation}
where $\Gamma_{1}$, $\Gamma_{2}$ and $\Gamma_{3}$ are three constants 
restricted by 
\begin{equation}
    \sum\limits_{i=1}^{3} \Gamma_{i} = 0
\label{restr:gamma}
\end{equation}
(we discuss this restriction in the conclusion).

The main object of this study are the `variational' equations, 
\begin{equation}
  \left. \partial \mathcal{S} \right/ \partial \myu(\vec{n}) = 0 
  \qquad
  (\myin{}) 
\end{equation}
which can be written as 
\begin{equation}
  \sum\limits_{i=1}^{3} 
  \Gamma_{i} 
  \left[  
    \frac{ \myu(\vec{n}+\vec{e}_{i}) }
         { 1 + \myu(\vec{n}) \myu(\vec{n}+\vec{e}_{i}) } 
    + 
    \frac{ \myu(\vec{n}-\vec{e}_{i}) }
         { 1 + \myu(\vec{n}) \myu(\vec{n}-\vec{e}_{i}) } 
  \right]
  = 0. 
\label{eq:main}
\end{equation}

In what follows, we extensively use the fact that the cubic lattice is a 
bipartite graph and split $\Lambda$ into the two sublattices, which we call 
`positive' and `negative': 
\begin{equation}
  \Lambda = \Lambda^{+} \cup \Lambda^{-}  
\end{equation}
where
\begin{equation}
 \begin{array}{l}
  \Lambda^{+} 
  = 
  \left\{ \left. 
    \vec{n} = \sum\limits_{i=1}^{3} n_{i} \vec{e}_{i}, 
    \quad
    n_{i} \in \mathbb{Z}
    \quad \right| \quad
    \sum\limits_{i=1}^{3} n_{i} = 0 \; \mbox{mod} \, 2 \; 
  \right\}, 
  \\[4mm] 
  \Lambda^{-} 
  = 
  \left\{ \left. 
    \vec{n} = \sum\limits_{i=1}^{3} n_{i} \vec{e}_{i}, 
    \quad
    n_{i} \in \mathbb{Z}
    \quad \right| \quad
    \sum\limits_{i=1}^{3} n_{i} = 1 \, \mbox{mod} \, 2 \; 
  \right\}. 
  \end{array}
\end{equation}
To compare the field equations of our model with already known systems, 
one can make the substitution 
\begin{equation} 
  \myu(\vec{n}) 
  = 
  \left\{
  \begin{array}{ccl} 
  \myw(\vec{n})    &\quad& (\myin{+}) \\
  -1/\myw(\vec{n}) && (\myin{-}) 
  \end{array}
  \right.
\end{equation}
which transforms equation \eref{eq:main} into 
\begin{equation}
  \sum\limits_{i=1}^{3} 
  \Gamma_{i} 
  \left[  
    \frac{ 1 }{ \myw(\vec{n}) - \myw(\vec{n}+\vec{e}_{i}) } 
    + 
    \frac{ 1 }{ \myw(\vec{n}) - \myw(\vec{n}-\vec{e}_{i}) } 
  \right]
  = 0. 
\label{eq:main-A}
\end{equation}
Written in this form, equation \eref{eq:main} can be viewed as a 
three-dimensional generalization of the already known integrable 
Toda-type and relativistic Toda-type lattices \cite{A00,A01,BS02,S03}. 
Indeed, in the case of $\Gamma_{1}=-\Gamma_{2}, \Gamma_{3}=0$
one arrives at the Toda-type lattice that belongs to the list of theorem 4 of 
\cite{A00} while another reduction of \eref{eq:main}, 
$\myw(\vec{n}+\vec{e}_{3}) = \myw(\vec{n}+\vec{e}_{1}+\vec{e}_{2})$, 
leads to the equation that belongs to the list of the lattices of the discrete 
relativistic Toda type of theorem 3 of \cite{A00}.

From the physical viewpoint, the action \eref{def:lnn} or \eref{def:model} 
describes an anharmonic lattice with the logarithmic interaction 
$V(\myu', \myu'') = \ln\left( 1 + \myu' \myu'' \right)$  
which \textit{i)} in the small amplitude limit becomes the `standard' harmonic 
one,  $V(\myu', \myu'') \approx \myu' \myu''$, 
and \textit{ii)} is not new to the theory of integrable systems. As a bright 
example of its appearance we should mention the classical$+$integrable analogue 
of the famous Heisenberg model of the quantum mechanics \cite{I82,H82,P87}). 
In this sense, the remarkable feature of equation \eref{eq:main} considered 
here is that it is one of a rather limited number of equations in 
\emph{multidimensions} that, on the one hand, are the field equation of a 
nonlinear lattice model with a reasonable action/energy and, on the other hand, 
possess (as is shown below) \emph{multi-soliton} solutions.

\section{Ansatz and bilinearization. \label{sec-bilin}}

In this section we present the main result of this paper. 
We bilinearize the field equation \eref{eq:main} and demonstrate the 
relationships of the resulting system with the already known 
integrable models. 

The aim of this paper is to derive the soliton solutions, i.e. 
some \emph{particular} solutions. To this end, we not only bilinearize the 
system \eref{eq:main} but also simplify it: we `split' the seven-point 
equations into four-point ones. 
The procedure that we use is, for the most part, rather standard. 
However, to achieve our goals we have to apply some non-trivial, 
though very simple, tricks. 

Below we discuss the main ideas behind the proposed substitutions and ansatz. 
Here we try to explain and `justify' the 
non-standard moments of the construction used in what follows.
A reader who considers that the only necessary justification of an ansatz is 
to check that it provides solutions for the equations in question 
may skip most of this section and 
proceed directly to proposition \ref{prop-bilin}.

\subsection{Three-leg reduction. }

The key idea behind the derivation of the soliton solutions for our model may 
be described as the three-leg reduction by analogy with the so-called 
three-leg representation of integrable systems on quad-graphs \cite{BS02} and 
which is known, for example, for all equations of the Adler-Bobenko-Suris list 
\cite{ABS03}. 
Using the language of \cite{BPS13}, we find the quad-equations for which 
action \eref{def:model} provides the so-called weak Lagrangian formulation.
The first manifestation of the three-dimensionality of our problem is that 
there is no clear way to build the system of polygons corresponding to our star 
equation. However a particular solution of this problem can be completed 
as follows. 

Consider the four vectors $\vec{g}_{\ell}$ ($\ell = 0,...,3$) given by 
\begin{equation}
  \vec{g}_{0} = \tfrac{1}{2}\sum_{i=1}^{3} \vec{e}_{i}, \qquad
  \vec{g}_{i} = \vec{e}_{i} - \vec{g}_{0} \quad (1 = 1,2,3) 
\label{def:vecg}
\end{equation}
related by 
\begin{equation}
  \sum_{\ell=0}^{3} \vec{g}_{\ell} = 0. 
\label{restr:vecg}
\end{equation}
Now, both 
$\myu(\vec{n}+\vec{e}_{i}) = \myu(\vec{n}+\vec{g}_{0}+\vec{g}_{i})$ 
and 
$\myu(\vec{n}-\vec{e}_{i}) = \myu(\vec{n}+\vec{g}_{j}+\vec{g}_{k})$ ($\myijk$) 
are obtained from $\myu(\vec{n})$ by means of two translations which surmise 
the following substitution: if we can find the function $\myv(\vec{n})$ such 
that 
\begin{equation}
\fl\quad
    \frac{ \myu(\vec{n}+\vec{g}_{l}+\vec{g}_{m}) }
         { 1 + \myu(\vec{n}) \myu(\vec{n}+\vec{g}_{l}+\vec{g}_{l}) } 
  = 
  \frac{ \myv(\vec{n} + \vec{g}_{l}) - \myv(\vec{n} + \vec{g}_{m}) } 
       { \mylambda{l} - \mylambda{m} },
  \quad l \ne m, \; l,m \in \{0,1,2,3\}, 
\label{quasi-ansatz}
\end{equation}
(the denominator in the right-hand side is introduced to preserve the symmetry 
with respect to the permutation $l \leftrightarrow m$)
then equation \eref{eq:main} can be written as 
\begin{equation}
  0 = 
  \sum_{i=1}^{3} 
  \hat\Gamma_{i} \; 
  \left[ \myv(\vec{n} + \vec{g}_{i}) - \myv(\vec{n} + \vec{g}_{0}) \right] 
\end{equation}
with \emph{constant} $\hat\Gamma_{i}$ and can be satisfied by making all 
$\hat\Gamma_{i}$ equal to zero (we return to this question in what follows). 

Of course, the function $\myv(\vec{n})$ is far from arbitrary: it has to meet 
various conditions following from the compatibility of the ansatz 
\eref{quasi-ansatz} together with its consistency with respect to translations.  
In particular, one can show that it must be a solution 
of certain Toda-type equation (we have some kind of duality here). 
So, to find suitable $\myv$ is not a trivial problem. However, one can hope to 
answer the arising questions in the framework of the \emph{four-point}, or 
quad-equations.
We do not discuss these problems now, because in what follows we will use, 
instead of ansatz \eref{quasi-ansatz}, another, more general,  one.

\subsection{Bipartite three-leg reduction. \label{sec:btlr} }

For our purposes, ansatz \eref{quasi-ansatz} has a serious drawback, 
stemming not from the compatibility/consistency issues, but from the 
restriction \eref{restr:vecg}. It turns out that when constructing explicit 
solutions such restrictions can be rather limiting. As we show in section 
\ref{sec:solitons}, they can leave us with only one- and two-soliton solutions. 
Thus, it seems useful to relieve us of the condition \eref{restr:vecg}, 
which can be done by elementary means. 

First, we replace the vectors 
$\{ \vec{g}_{\ell} \}_{\ell=0}^{3}$ with another set of vectors, 
$\{ \vec{\alpha}_{\ell} \}_{\ell=0}^{3}$, that are related to 
$\{ \vec{e}_{i} \}_{i=1}^{3}$ by 
\begin{equation}
  \vec{e}_{i} 
  = 
  \tfrac{1}{2} 
  \left( 
    \vec{\alpha}_{0} + \vec{\alpha}_{i} 
    - \vec{\alpha}_{j} - \vec{\alpha}_{k} 
  \right). 
\label{def:alpha}
\end{equation}
It is obvious that one can find infinitely many such quadruples. 
For example, one can 
take arbitrary $\vec{\alpha}_{0}$ and then put 
$\vec{\alpha}_{i} = \vec{\alpha}_{0} - \vec{e}_{j} - \vec{e}_{k}$ 
($\myijk$). 
However, in our case this ambiguity is not a problem. 
As is shown below, these vectors appear in solutions that we are going to 
derive as parameters, hence different choices of 
$\{ \vec{\alpha}_{\ell} \}_{\ell=0}^{3}$ just provide, in general, different 
solutions. 

Now, one can present $\myu(\vec{n} \pm \vec{e}_{i})$ as  
\begin{equation}
  \begin{array}{lcl}
  \myu(\vec{n} + \vec{e}_{i}) 
  & = & 
  \myu(\vec{n} + \vec{\alpha}_{0} + \vec{\alpha}_{i} - 2\vec{\delta} )	
  \\ 
  \myu(\vec{n} - \vec{e}_{i}) 
  & = & 
  \myu(\vec{n} + \vec{\alpha}_{j} + \vec{\alpha}_{k} - 2\vec{\delta} )	
  \end{array}
  \qquad\myijk 
\end{equation} 
where 
\begin{equation}
  \vec{\delta} 
  = 
  \tfrac{1}{4} \sum_{\ell=0}^{3} \vec{\alpha}_{\ell}, 
\end{equation}
i.e. the translations by the vectors $\pm\vec{e}_{i}$ cease to be sums of two 
translations, the fact that has been crucial for the ansatz 
\eref{quasi-ansatz}. 
To restore this feature of   
$\{ \vec{\alpha}_{\ell} \}_{\ell=0}^{3}$, 
we use the bipartite property of the cubic lattice and introduce, instead of 
$\myu$, two functions, $q$ and $r$, by  
\begin{equation} 
  \myu(\vec{n}) 
  = 
  \left\{
  \begin{array}{lcl} 
  r(\vec{n} - \vec{\delta}) &\quad& (\myin{+}) \\
  q(\vec{n} + \vec{\delta}) && (\myin{-}) 
  \end{array}
  \right.
\label{def:qr}
\end{equation}
In terms of $q$ and $r$ equation \eref{eq:main} becomes  
\begin{equation}
\label{bi:eq}
\fl
  \begin{array}{l}
  \displaystyle 
  0 = 
  \sum_{i=1}^{3} 
  \Gamma_{i} \; 
  \left[ 
    \frac{ q(\vec{x}_{+} + \vec{\alpha}_{0} + \vec{\alpha}_{i}) } 
         { 1 +  
           q(\vec{x}_{+} + \vec{\alpha}_{0} + \vec{\alpha}_{i}) 
           r(\vec{x}_{+}) } 
    + 
    \frac{ q(\vec{x}_{+} + \vec{\alpha}_{j} + \vec{\alpha}_{k}) } 
         { 1 +  
           q(\vec{x}_{+} + \vec{\alpha}_{j} + \vec{\alpha}_{k}) 
           r(\vec{x}_{+}) } 
  \right] 
  \; (\myin{+})
  \\[4mm]
  \displaystyle 
  0 = 
  \sum_{i=1}^{3} 
  \Gamma_{i} \; 
  \left[ 
    \frac{ r(\vec{x}_{-} - \vec{\alpha}_{0} - \vec{\alpha}_{i}) } 
         { 1 +  
           q(\vec{x}_{-}) 
           r(\vec{x}_{-} - \vec{\alpha}_{0} - \vec{\alpha}_{i}) } 
    + 
    \frac{ r(\vec{x}_{-} - \vec{\alpha}_{j} - \vec{\alpha}_{k}) } 
         { 1 +  
           q(\vec{x}_{-}) 
           r(\vec{x}_{-} - \vec{\alpha}_{j} - \vec{\alpha}_{k}) } 
  \right] 
  \; (\myin{-}) 
  \end{array}
\end{equation}
where $ \vec{x}_{\pm} = \vec{n} \mp \vec{\delta}$. 

The original problem is \emph{discrete}: all functions were defined on 
$\mathbb{Z}^{3}$. 
If we were using \eref{quasi-ansatz}, we would actually pass to the 
body-centered cubic lattice, but the problem would remain discrete. 
However, now, after introducing the one-parametric family of vectors 
$\vec\alpha$ and functions $q$ and $r$ that have different domains of 
definition, we make next step and consider the above equations as defined on 
the whole $\mathbb{R}^{3}$, 
\begin{equation}
\label{bi:qr} 
\fl
  \left\{ 
  \begin{array}{l} 
  \displaystyle 
	0 = 
  \sum_{i=1}^{3} 
  \Gamma_{i} \; 
  \left[ 
    \frac{ q(\vec{x} + \vec{\alpha}_{0} + \vec{\alpha}_{i}) } 
         { 1 +  
           q(\vec{x} + \vec{\alpha}_{0} + \vec{\alpha}_{i}) 
           r(\vec{x}) } 
    + 
    \frac{ q(\vec{x} + \vec{\alpha}_{j} + \vec{\alpha}_{k}) } 
         { 1 +  
           q(\vec{x} + \vec{\alpha}_{j} + \vec{\alpha}_{k}) 
           r(\vec{x}) } 
  \right] 
  \\[4mm] \displaystyle 
  0 = 
  \sum_{i=1}^{3} 
  \Gamma_{i} \; 
  \left[ 
    \frac{ r(\vec{x} - \vec{\alpha}_{0} - \vec{\alpha}_{i}) } 
         { 1 +  
           q(\vec{x}) 
           r(\vec{x} - \vec{\alpha}_{0} - \vec{\alpha}_{i}) } 
    + 
    \frac{ r(\vec{x} - \vec{\alpha}_{j} - \vec{\alpha}_{k}) } 
         { 1 +  
           q(\vec{x}) 
           r(\vec{x} - \vec{\alpha}_{j} - \vec{\alpha}_{k}) } 
  \right] 
  \end{array}
  \right.   
  \; 
  (\vec{x} \in \mathbb{R}^{3}). 
\end{equation}
In other words, we pass from discrete equations to difference ones, which is, 
of course a reduction. However this reduction is rather typical for the 
`applied' studies of discrete equations. 
Moreover, we repeat that we are looking for some particular solutions and 
hence can admit some reductions, provided they leave us with non-trivial 
residue. 
Looking at soliton solutions for various models one can see that, usually, they 
depend analytically on all arguments and parameters and can be obtained as 
solutions of corresponding difference equations.
Finally, this question will be less significant when we finish the derivation 
of the soliton solutions. The final formulae are written in terms of 
$\vec{n}$ with $\vec{\alpha}$ being replaced with 
corresponding parameters and can be considered as solutions for the pure 
discrete problem.

To derive solutions for system \eref{bi:qr}, we make the following ansatz: 
\begin{equation}
\label{bi:ansatz}
  \begin{array}{l} 
  \displaystyle
  \frac{ q(\vec{x} + \vec{\alpha}_{l} + \vec{\alpha}_{m}) } 
       { 1 +  
         q(\vec{x} + \vec{\alpha}_{l} + \vec{\alpha}_{m}) r(\vec{x}) } 
  = 
  \frac{ q(\vec{x} + \vec{\alpha}_{l}) - q(\vec{x} + \vec{\alpha}_{m}) } 
       { \mylambda{l} - \mylambda{m} } 
  \\[4mm] 
  \displaystyle
  \frac{ r(\vec{x} - \vec{\alpha}_{l} - \vec{\alpha}_{m}) }
       { 1 +  q(\vec{x}) r(\vec{x} - \vec{\alpha}_{l} - \vec{\alpha}_{m}) } 
  = 
  \frac{ r(\vec{x} - \vec{\alpha}_{l}) - r(\vec{x} - \vec{\alpha}_{m}) } 
       { \mylambda{l} - \mylambda{m} },
  \end{array}
  \quad 
\end{equation}
for $l,m \in \{0,1,2,3\}$ and $l \ne m$ (the simplest, self-dual, form of 
equation \eref{quasi-ansatz}, discussed in the previous subsection), which, as 
is shown below, reduces the problem to the already known system (which also 
helps us to answer the questions about its compatibility) and, what is 
important for this work, leads to the N-soliton solutions.

Ansatz \eref{bi:ansatz} reduces \eref{bi:qr} to 
\begin{equation}
  \left\{
  \begin{array}{l}
  0 = 
  \sum_{i=1}^{3} 
  \hat\Gamma_{i} \; 
  \left[ q(\vec{x} + \vec{\alpha}_{i}) - q(\vec{x} + \vec{\alpha}_{0}) \right] 
  \\[2mm] 
  0 = 
  \sum_{i=1}^{3} 
  \hat\Gamma_{i} \; 
  \left[ r(\vec{x} - \vec{\alpha}_{i}) - r(\vec{x} - \vec{\alpha}_{0}) \right] 
  \end{array} 
  \right.
\label{bi:ansatz-result}
\end{equation}
with constants $\hat\Gamma_{i}$ given by 
\begin{equation}
  \hat\Gamma_{i} 
  = 
  \frac{ \Gamma_{i} }
       { \mylambda{i} - \mylambda{0} } 
  + 
  \frac{ \Gamma_{j} }
       { \mylambda{i} - \mylambda{k} } 
  +
  \frac{ \Gamma_{k} }
       { \mylambda{i} - \mylambda{j} } 
  \qquad \myijk. 
\label{def:hat-Gamma}
\end{equation}

It easy to see that we can satisfy equations \eref{bi:ansatz-result} without 
imposing additional conditions upon $q$ and $r$ by making all $\hat\Gamma_{i}$ 
equal to zero. 
Solution of this elementary problem leads to the following restriction on 
the constants $\mylambda{\ell}$: 
\begin{equation}
  \mylambda{0} 
  = 
  \sum_{i=1}^{3} \mylambda{i} 
  - 
  \frac{ \sum_{i=1}^{3} \Gamma_{i} \mylambda[2]{i} } 
       { \sum_{i=1}^{3} \Gamma_{i} \mylambda{i} } 
\label{restr:lambda}
\end{equation}
(see \ref{app:lambda} for a proof). 

\subsection{Bilinearization. }

Finally, we bilinearize the system \eref{bi:ansatz}
by introducing the triplet of the tau-functions $\sigma$, $\rho$ and $\tau$: 
\begin{equation}
  q = \frac{ \sigma }{ \tau },
  \qquad
  r = \frac{ \rho }{ \tau }. 
\label{def:tau}
\end{equation}
It is easy to check that $q$ and $r$ are solutions for \eref{bi:ansatz} 
provided $\sigma$, $\rho$ and $\tau$ satisfy 
\mybeginsubeqs{syst:hal}
\begin{eqnarray}
\fl\qquad
&&
  \myconst{a}{l,m}  
  \tau \, \sigma(\vec{x} + \vec{\alpha}_{l} + \vec{\alpha}_{m}) 
  = 
  \sigma(\vec{x} + \vec{\alpha}_{l}) \, 
  \tau(\vec{x} + \vec{\alpha}_{m})
  - 
  \tau(\vec{x} + \vec{\alpha}_{l}) \, 
  \sigma(\vec{x} + \vec{\alpha}_{m}) 
\label{syst:hal-s}
\\[1mm]
\fl\qquad
&&
  \myconst{a}{l,m}  
  \rho \, \tau(\vec{x} + \vec{\alpha}_{l} + \vec{\alpha}_{m}) 
  = 
  \tau(\vec{x} + \vec{\alpha}_{l}) \, 
  \rho(\vec{x} + \vec{\alpha}_{m}) 
  - 
  \rho(\vec{x} + \vec{\alpha}_{l}) \, 
  \tau(\vec{x} + \vec{\alpha}_{m}) 
\label{syst:hal-r}
\\[1mm]
\fl\qquad
&&
  \myconst{b}{l,m}
  \tau(\vec{x} + \vec{\alpha}_{l}) \, 
  \tau(\vec{x} + \vec{\alpha}_{m}) 
  = 
  \tau \, \tau(\vec{x} + \vec{\alpha}_{l} + \vec{\alpha}_{m}) 
  + 
  \rho \, \sigma(\vec{x} + \vec{\alpha}_{l} + \vec{\alpha}_{m})  
\label{syst:hal-t}
\end{eqnarray}
\myendsubeqs
where $l,m \in \{0,1,2,3\}$ and $l \ne m$, 
the skew-symmetric constants $\myconst{a}{l,m}$ and symmetric 
constants $\myconst{b}{l,m}$  are related to $\mylambda{\ell}$ by
\begin{equation}
  \myconst{a}{l,m}
  \myconst{b}{l,m}
  =
  \mylambda{l} - \mylambda{m} 
\label{bi:abll}
\end{equation}
but are arbitrary apart from that.

To summarize, the main result of this paper can be presented as follows. 
%
\begin{proposition} \label{prop-bilin}
Each solution of system \eref{syst:hal} delivers a solution for the field 
equation \eref{eq:main}, by means of \eref{def:qr} and \eref{def:tau}, 
provided the parameters $\{ \lambda_{l} \}_{\ell=0}^{3}$ 
satisfy restrictions \eref{restr:lambda} and \eref{bi:abll}.

\end{proposition} 

The proof of this statement is straightforward:
system \eref{syst:hal}, together with \eref{bi:abll}, implies that the functions 
$q$ and $r$ given by \eref{def:tau} satisfy \eref{bi:qr}. As is shown in 
section \ref{sec:btlr}, this and \eref{restr:lambda} guarantee that the 
functions $\myu$ given by \eref{def:qr} solve \eref{eq:main}. 

\subsection{Hirota-Ablowitz-Ladik system. \label{sec-alh}} 

The system \eref{syst:hal} is an already known system that can be found 
in studies of a large number of integrable equations 
(see, e.g., \cite{H81,DJM83,SS87,PT07,WH15,M15}). 
Probably, the most important appearance of \eref{syst:hal} 
is in the theory of such well-studied integrable models as the HBDE \cite{H81} 
and the ALH \cite{AL75}. 
For example, an immediate consequence of \eref{syst:hal-s}, or 
\eref{syst:hal-r}, is the fact that $\tau$ solves the HBDE, 
so both \eref{syst:hal-s} and \eref{syst:hal-r} can be viewed as linear 
problems from the zero-curvature representation of the HBDE. 
On the other hand, equations \eref{syst:hal-s} and \eref{syst:hal-r} describe 
the infinite chain of the B\"acklund transformations for the HBDE. 
It is also known that system \eref{syst:hal} is closely related to another 
integrable model, which is even `older' than the HBDE: equations 
\eref{syst:hal} describe the so-called Miwa shifts of the ALH. 
We do not discuss these questions here in detail referring the reader to 
section 4.1 (together with appendices A and B) of \cite{V16a} and section 4 of 
\cite{V16b}.  

The fact that we have reduced our  problem to the well-studied system 
\eref{syst:hal} has two advantages. First, we do not need to worry about the 
compatibility of the ansatz \eref{bi:ansatz} or its consistency 
(see, e.g., \cite{PT07}). 
Second, we can use the wide number of solutions already derived for 
\eref{syst:hal}. In this paper, we discuss the soliton ones (see the following 
section). However, we might obtain by the same ansatz the so-called finite-gap 
quasi-periodic solutions, for which system \eref{syst:hal} is just the set of 
Fay identities, or various determinant solutions.

\section{Soliton solutions. \label{sec:solitons}} 

In what follows, we derive soliton solutions for our problem using the results 
of \cite{V14,V15}, where we have presented a large number of identities 
for the matrices of a special type (soliton Fay identities).

In papers \cite{V14,V15} we describe two types of constructions that lead to 
the soliton solutions for various models. In the case of one spatial dimension 
the difference between these solitons usually indicated by the words `bright' 
and `dark': bright solitons satisfy the zero boundary conditions while the dark 
ones (or their absolute values) tend to constants. In the multidimensional case, 
the situation is more complicated. For example, the simplest analogues of the 
one-dimensional bright solitons (one-soliton solutions) in multidimensions 
become the line solitons which decay in one direction (and its opposite) 
but are constant in orthogonal ones. 
Thus, `bright' and `dark' are not the best terms to classify the solitons in 
multidimensional models as the one of this 
paper. However, we will use them in what follows, just to distinguish the two 
types of solutions: 
solutions constructed from the matrices described in \cite{V14}, that in 
a one-dimensional case lead to the dark solitons, 
and 
ones constructed from the matrices described in \cite{V15}, that in 
a one-dimensional case lead to the bright solitons.

\subsection{`Dark' solitons.} 

Here we use one of the results of \cite{V14} which (after some simplifications) 
can be formulated as follows: the determinants 
\begin{equation}
  \Omega = 
  \det\left| \mymatrix{1} + \mymatrix{A} \right|
\end{equation}
of the matrices defined by 
\begin{equation}
  \mymatrix{L} \mymatrix{A} - \mymatrix{A} \mymatrix{L}^{-1} 
  = 
  | \, 1 \, \rangle \langle a |, 
\label{dark:LA}
\end{equation}
where 
$
  \mymatrix{L} = \mbox{diag}\left( L_{1}, ..., L_{N} \right), 
$
$| \, 1 \, \rangle$ is the $N$-column with all components equal to $1$, 
$\langle a |$ is a $N$-component row that depends on the 
coordinates describing the model, 
satisfy 
\begin{equation}
\fl\qquad
  0 = 
  (\xi - \eta) \myShifted[1]{\xi\eta}{\Omega} \myShifted[1]{\zeta}{\Omega} 
  + 
  (\zeta - \xi) \myShifted[1]{\xi\zeta}{\Omega} \myShifted[1]{\eta}{\Omega} 
  + 
  (\eta - \zeta) \myShifted[1]{\eta\zeta}{\Omega} \myShifted[1]{\xi}{\Omega}. 
\label{dark:fay}
\end{equation}
Here, the shifts $\myShift{}$ are defined as 
$
  \myShifted{\xi}{\Omega} = 
  \det\left| \mymatrix{1} + \myShifted{\xi}{\mymatrix{A}} \right|, 
$
$\myShifted{\xi\eta}{\Omega} = \myShift{\xi}\!\myShift{\eta}\Omega$ 
with 
\begin{equation}
  \myShifted{\xi}{\mymatrix{A}} 
  = 
  \mymatrix{A}\mymatrix{H}_{\xi} 
\end{equation}
and 
\begin{equation}
  \mymatrix{H}_{\xi} 
  = 
  \left( \xi - \mymatrix{L} \right) 
  \left( \xi - \mymatrix{L}^{-1} \right)^{-1}. 
\end{equation}
From this identity, together with the `duality' property of the matrices 
$\mymatrix{H}_{\xi}$,  
\begin{equation}
  \mymatrix{H}_{\xi} 
  \mymatrix{H}_{1/\xi} 
  = 
  \mymatrix{H}_{0}, 
\end{equation}
which implies 
\begin{equation}
  \myShift{\xi} \myShift{1/\xi} 
  = 
  \myShift{0} 
\label{dark:duality}
\end{equation}
it is easy to derive
%
\begin{propositionA} 
\label{dark:prop-hal}
Functions 
\begin{equation}
  \tau = \Omega, 
  \quad
  \sigma = - F^{-1} \, \myShifted[2]{0}{\Omega}, 
  \quad
  \rho = F \myShifted{0}{\Omega}
\end{equation}
where $F$ is defined by 
\begin{equation}
  \myShifted{\alpha}{F} = - \alpha F
\end{equation}
satisfy the system  
\mybeginsubeqs{syst:alh}
\begin{eqnarray}
&&
  \myalhconst{a}{\xi,\eta}
  \tau
  \myShifted[1]{\xi\eta}{\sigma}
  = 
  \myShifted[1]{\xi}{\sigma} \myShifted[1]{\eta}{\tau}
  - 
  \myShifted[1]{\xi}{\tau} \myShifted[1]{\eta}{\sigma}, 
\label{syst:alh-s}
\\[1mm]&&
  \myalhconst{a}{\xi,\eta}
  \rho
  \myShifted[1]{\xi\eta}{\tau}
  = 
  \myShifted[1]{\xi}{\tau} \myShifted[1]{\eta}{\rho}
  - 
  \myShifted[1]{\xi}{\rho} \myShifted[1]{\eta}{\tau}, 
\label{syst:alh-r}
\\[1mm]&&
  \myalhconst{b}{\xi,\eta}
  \myShifted[1]{\xi}{\tau}
  \myShifted[1]{\eta}{\tau}
  = 
  \tau \myShifted[1]{\xi\eta}{\tau}
  + 
  \rho \myShifted[1]{\xi\eta}{\sigma} 
\label{syst:alh-t}
\end{eqnarray}
\myendsubeqs
with 
\begin{equation}
  \myalhconst{a}{\alpha,\beta}
  = 
  \alpha - \beta, 
\qquad
  \myalhconst{b}{\alpha,\beta}
  = 
  1 - \frac{ 1 }{\alpha\beta}. 
\label{dark:ab} 
\end{equation}

\end{propositionA} 
\medskip\noindent
(see \ref{app:dark} for a proof).

It is clear that \eref{syst:alh} is exactly \eref{syst:hal}, provided 
we identify the translations $\vec{x} \to \vec{x} + \vec{\alpha}_{\ell}$ with 
the action of $\myShift{\alpha_{\ell}}$, 
where $\{ \alpha_{\ell} \}_{\ell=0}^{3}$ is a fixed set of constants, 
and put 
$\myconst{a}{l,m} = \myalhconst{a}{\alpha_{l},\alpha_{m}}$ 
and 
$\myconst{b}{l,m} = \myalhconst{b}{\alpha_{l},\alpha_{m}}$, 
which implies 
\begin{equation}
  \mylambda{\ell}
  = 
  \alpha_{\ell} + \alpha_{\ell}^{-1} 
\end{equation}
(up to a non-essential constant). 

Thus, we have all that is necessary to write down the `dark'-soliton 
solutions for the field equations of our model.

Using the construction described in section \ref{sec:btlr}, we write the 
relation between the translations by $\vec{e}_{i}$ and the action of the 
shifts $\myShift{\alpha_{\ell}}$ as 
\begin{equation}
  \myu(\vec{n} + \vec{e}_{i}) 
  = 
  \myShift{\alpha_{0}}^{1/2}
  \myShift{\alpha_{i}}^{1/2}
  \myShift{\alpha_{j}}^{-1/2}
  \myShift{\alpha_{k}}^{-1/2}
  \myu(\vec{n}) 
  \qquad \myijk 
\label{def:translation}
\end{equation}
or 
\begin{equation}
  \myu(\vec{n} + \vec{e}_{i}) 
  = 
  \myShift{\alpha_{i}} \myShift{*}^{-1} \myu(\vec{n}), 
  \qquad
  \myShift{*} 
  = 
  \left[ 
    \myShift{\alpha_{0}}^{-1} 
    \myShift{\alpha_{1}} 
    \myShift{\alpha_{2}} 
    \myShift{\alpha_{3}} 
  \right]^{1/2}.
\end{equation}
Then, we introduce the matrices $\mymatrix{X}_{i}$ ($i=1,2,3$) 
\begin{equation}
\label{dark:Xi}
  \mymatrix{X}_{i} 
  = 
  \mymatrix{H}_{\alpha_{i}} 
  \mymatrix{H}_{*}^{-1}, 
\qquad
  \mymatrix{H}_{*} 
  = 
  \left[ 
    \mymatrix{H}_{\alpha_{0}}^{-1}  
    \mymatrix{H}_{\alpha_{1}} 
    \mymatrix{H}_{\alpha_{2}} 
    \mymatrix{H}_{\alpha_{3}} 
  \right]^{1/2}   
\end{equation}
describing the $\vec{n}$-dependence, 
\begin{equation} 
  \mymatrix{A}\left( \vec{n} + \vec{e}_{i} \right) 
  = 
  \mymatrix{A}\left( \vec{n} \right) \mymatrix{X}_{i}
\end{equation}
as well as two matrices, $\mymatrix{M}_{0}$ and $\mymatrix{M}_{1}$ which 
describe the action of 
$\myShift{\delta}^{-1}$ and $\myShift{0} \myShift\delta^{-1}$, where 
$\myShift{\delta}$ is the shift corresponding to the translation 
$\vec{x} \to \vec{x} + \vec{\delta}$, 
$
  \myShift{\delta}  
  = 
  \left( \prod\nolimits_{\ell=0}^{3} \myShift{\alpha_{\ell}} \right)^{1/4}
$, 
\begin{equation}
  \mymatrix{M}_{0} 
  = 
  \left.\overline{\mymatrix{H}}\right.^{-1}, 
  \qquad
  \mymatrix{M}_{1} 
  = 
  \mymatrix{L}^{2} \left.\overline{\mymatrix{H}}\right.^{-1} 
\end{equation}
where
\begin{equation}
  \overline{\mymatrix{H}} 
  = 
  \left[ \prod\nolimits_{\ell=0}^{3} \mymatrix{H}_{\alpha_{\ell}} \right]^{1/4}. 
\label{dark:barH}
\end{equation}

It should be noted that calculating the action of the shift 
$\myShift\delta$ on the matrices $\mymatrix{A}$ and the function $F$ one has 
to raise the products 
$\prod\nolimits_{\ell=0}^{3} \mymatrix{H}_{\alpha_{\ell}}$ 
and
$\prod\nolimits_{\ell=0}^{3} \alpha_{\ell}$ to the power $1/4$. 
This leads to some restrictions on the parameters $\alpha_{\ell}$ and 
$L_{n}$:
\mybeginsubeqs{dark:restr}
\begin{equation}
  \prod\nolimits_{\ell=0}^{3} \alpha_{\ell} > 0, 
\end{equation}
which should be considered together with equation \eref{restr:lambda} and 
\begin{equation}
  \prod\nolimits_{\ell=0}^{3} 
  \left[ 
    L_{n} + L_{n}^{-1} 
    - 
    \left( \alpha_{\ell} + \alpha_{\ell}^{-1} \right) 
  \right] 
  > 0 
  \qquad
  ( n = 1, ..., N)
\end{equation}
\myendsubeqs 

As the result we can formulate 
%
\begin{propositionA} 
The `dark'-soliton solutions can be presented as 
\begin{equation}
  \myu(\vec{n}) 
  =
  \pm \, u_{*} \, F(\vec{n})^{\pm 1} \; 
  \frac{
    \det\left| 
    \mymatrix{1} 
    + 
    \mymatrix{A}(\vec{n}) \, \mymatrix{M}_{1}^{\pm 1} 
    \right| 
  }
  {
    \det\left| 
    \mymatrix{1} 
    + 
    \mymatrix{A}(\vec{n}) \, \mymatrix{M}_{0}^{\pm 1} 
    \right| 
  }
\qquad
  (\myin{\pm})
\end{equation}
where
$u_{*} = \left( \alpha_{0} \alpha_{1} \alpha_{2} \alpha_{3} \right)^{-1/4} $,
\begin{equation}
\label{dark:F}
  F(\vec{n}) 
  = 
  \prod_{i=1}^{3} (\alpha_{i}/\alpha_{*})^{n_{i}}, 
  \qquad 
  \alpha_{*} 
  = 
  \left( \alpha_{1}\alpha_{2}\alpha_{3} / \alpha_{0} \right)^{1/2}, 
\end{equation}
\begin{equation}
  \mymatrix{A}(\vec{n}) 
  = 
  \mymatrix{C} \, 
  \mymatrix{X}(\vec{n}), 
  \qquad 
  \mymatrix{X}(\vec{n}) =   
  \prod_{i=1}^{3} \mymatrix{X}_{i}^{n_{i}} 
\end{equation}
($n_{i}$ are the components of $\vec{n}$, 
$\vec{n}=\sum_{i=1}^{3} n_{i} \vec{e}_{i}$), 
with the matrices $\mymatrix{X}_{i}$ being given by \eref{dark:Xi} 
and
\begin{equation}
  \mymatrix{C} 
  = 
  \left( 
    \frac{ c_{n} }{ L_{m}L_{n} - 1 } 
  \right)_{m,n=1, ..., N}. 
\end{equation}
Here, 
$\alpha_{0} = \alpha_{0}\left(\alpha_{1},\alpha_{2},\alpha_{3}\right)$  
is a solution of 
\begin{equation}
  \alpha_{0} + \alpha_{0}^{-1} 
  = 
  \sum_{i=1}^{3} ( \alpha_{i} + \alpha_{i}^{-1} ) 
  - 
  \frac{ \sum_{i=1}^{3} \Gamma_{i} ( \alpha_{i}^{2} + \alpha_{i}^{-2} ) } 
       { \sum_{i=1}^{3} \Gamma_{i} ( \alpha_{i} + \alpha_{i}^{-1} ) } 
\end{equation}
and 
$c_{n}$, $L_{n}$ ($n=1, ..., N$) and 
$\alpha_{i}$ ($i=1,2,3$) are arbitrary (up to the restrictions 
\eref{dark:restr}) constants.

\end{propositionA} 

The simplest, 1-soliton solution can be rewritten as 
\begin{equation}
  \myu(\vec{n}) 
  =
  \pm \, u_{*} \, 
  e^{ \pm f(\vec{n}) } \; 
  \frac{ \cosh\left( h(\vec{n}) \pm \delta_{1} \right) } 
       { \cosh\left( h(\vec{n}) \pm \delta_{0} \right) } 
\qquad
  (\myin{\pm})
\end{equation}
where 
$u_{*} = 1 / \bar\alpha$, 
\begin{equation}
  f(\vec{n}) = f_{0} + \left( \vec{\varphi}, \vec{n} \right), 
  \qquad 
  h(\vec{n}) = h_{0} + \left( \vec{\chi}, \vec{n} \right) 
\end{equation}
with arbitrary $f_{0}$ and $h_{0}$ and 
\begin{equation}
  \begin{array}{lcl}
  \vec{\varphi} =  \sum_{i=1}^{3} \varphi_{i} \vec{e}_{i}, 
  &&
  \varphi_{i}   = \ln \,\frac{ \alpha_{i} }{ \alpha_{*} }, 
  \\[2mm]
  \vec{\chi}    = \sum_{i=1}^{3} \chi_{i} \vec{e}_{i}, 
  &&
  \chi_{i}      = \frac{1}{2} \ln \frac{ H_{i} }{ H_{*} }. 
  \end{array}
\end{equation}
Here, $H_{i} = (\alpha_{i} - L)/(\alpha_{i} - 1/L)$, and $L$ are scalars 
replacing the matrices $\mymatrix{H}_{\alpha_{i}}$ and $\mymatrix{L}$, 
instead of the matrices $\mymatrix{M}_{0,1}$ we use the constants 
$\delta_{0,1}$ given by, 
\begin{equation}
  \delta_{0} = - \tfrac{1}{2} \ln \bar{H}, 
  \qquad
  \delta_{1} = \tfrac{1}{2} \ln\left( L^{2} / \bar{H} \right) 
\end{equation}
with 
\begin{equation}
  \bar{\alpha} = \left( \alpha_{0}\alpha_{1}\alpha_{2}\alpha_{3} \right)^{1/4}, 
  \qquad
  \bar{H} = \left( H_{0}H_{1}H_{2}H_{3} \right)^{1/4} 
\end{equation}
and  
\begin{equation}
  \alpha_{*} = 
  \left( \alpha_{1}\alpha_{2}\alpha_{3} / \alpha_{0} \right)^{1/2}, 
  \qquad
  H_{*} = 
  \left( H_{1}H_{2}H_{3} / H_{0} \right)^{1/2}. 
\end{equation}
It is easy to see that we have a real analogue of the dark soliton of the 
complex models like the nonlinear Schr\"{o}dinger or the Ablowitz-Ladik 
equations: the plane wave $\pm \, u_{*} \, e^{ \pm f(\vec{n}) }$ 
(in the complex case $f(\vec{n})$ is pure imaginary) 
modulated by the factor which in the $\left| \vec{n} \right| \to \infty$ 
limit tends to one of the two constants, $L$ or $1/L$, depending on 
the direction $\vec{n}/\left|\vec{n}\right|$ in which we approach the infinity
(that determines the sign of $\left( \vec{\chi}, \vec{n} \right)$).

\subsection{`Bright' solitons.} 

To derive the second type of soliton solutions one does not need any additional 
calculations but can use the soliton Fay identities from \cite{V15} which were 
obtained for the tau-functions
\begin{equation}
  \begin{array}{lcl}
  \tau & = &\det | \mymatrix{1} + \mymatrix{A}\mymatrix{B} |, 
  \\[1mm] 
  \sigma 
  & = & 
  \tau \langle a | 
  ( \mymatrix{1} + \mymatrix{B}\mymatrix{A} )^{-1} 
  |1\rangle, 
  \\[1mm] 
  \rho 
  & = & 
  \tau \langle b | 
  ( \mymatrix{1} + \mymatrix{A}\mymatrix{B} )^{-1} 
  |1\rangle  
  \end{array} 
\end{equation}
where $\mymatrix{A}$ and $\mymatrix{B}$ are solutions of 
\begin{equation}
  \begin{array}{lcl}
  \mymatrix{L} \mymatrix{A} - \mymatrix{A} \mymatrix{R} 
  & = & 
  |1\rangle \langle a |, 
  \\[1mm] 
  \mymatrix{R} \mymatrix{B} - \mymatrix{B} \mymatrix{L} 
  & = & 
  |1\rangle \langle b |. 
  \end{array}
\end{equation}
Here, like in the previous section, 
$\mymatrix{L}$ and $\mymatrix{R}$ are constant diagonal $N \times N$-matrices, 
$\mymatrix{L} = \mbox{diag}\left( L_{1}, ..., L_{N} \right)$ and
$\mymatrix{R} = \mbox{diag}\left( R_{1}, ..., R_{N} \right)$, 
$| \, 1 \, \rangle$ is the $N$-column with all components equal to $1$, 
$\langle a |$ and $\langle b |$ are $N$-component rows that depend on the 
coordinates describing the model. 

The shifts $\myShift{\xi}$ are defined, in this case, by 
\begin{equation}
  \begin{array}{lcl}
  \myShifted{\xi}{\langle a |} 
  & = & 
  \langle a | \left(\mymatrix{R} - \xi\right)^{-1}, 
  \\[1mm]
  \myShifted{\xi}{\langle b |} 
  & = & 
  \langle b | \left(\mymatrix{L} - \xi\right) 
  \end{array} 
\end{equation}
or, as a consequence, by 
\begin{equation}
  \begin{array}{lcl}
  \myShifted{\xi}{\mymatrix{A}} 
  & = & 
  \mymatrix{A} \left(\mymatrix{R} - \xi\right)^{-1}, 
  \\[1mm]
  \myShifted{\xi}{\mymatrix{B}} 
  & = & 
  \mymatrix{B} \left(\mymatrix{L} - \xi\right). 
  \end{array} 
\end{equation}
The simplest soliton Fay identities, which are equations (3.12)--(3.14) of 
\cite{V15}, are exactly equations \eref{syst:alh} with 
\begin{equation}
  \myalhconst{a}{\xi,\eta} = \xi-\eta, 
  \quad
  \myalhconst{b}{\xi,\eta}=1 
\label{bright:ab}
\end{equation}
which implies 
\begin{equation}
  \mylambda{\ell} = \alpha_{\ell}. 
\end{equation}
Thus, the only thing that we have to do to obtain solutions for our equations 
is to gather all formulae describing $\myu(\vec{n})$ in terms of the 
tau-functions and the shifts $\myShift{\alpha_{\ell}}$, 
with a fixed set of constants $\{ \alpha_{\ell} \}_{\ell=0}^{3}$, 
which correspond, as in the previous subsection, 
to the translations $\vec{x} \to \vec{x} + \vec{\alpha}_{\ell}$.

To make the final formulae more clear we change the notation: we write 
$\mymatrix{L}^{\pm}$ instead of $\mymatrix{L}$ and $\mymatrix{R}$, 
\begin{equation}
  \mymatrix{L}^{+} = \mymatrix{R}, 
  \qquad
  \mymatrix{L}^{-} = \mymatrix{L}, 
\end{equation}
and slightly modify the definition of the rows $\langle a |$ and $\langle b |$ 
and of the matrices $\mymatrix{A}$ and $\mymatrix{B}$: 
we use in what follows the new rows  
\begin{equation}
  \langle a^{+} | = \langle b | {\overline{\mymatrix{L}}\,}^{-1}, 
  \qquad 
  \langle a^{-} | = \langle a | {\overline{\mymatrix{R}}\,}^{-1}   
\end{equation}
and the new matrices 
\begin{equation}
  \mymatrix{A}^{+} = \mymatrix{B} {\overline{\mymatrix{L}}\,}^{-1}, 
  \qquad 
  \mymatrix{A}^{-} = \mymatrix{A} {\overline{\mymatrix{R}}\,}^{-1} 
\end{equation}
where $\overline{\mymatrix{L}}$ and $\overline{\mymatrix{R}}$ are 
the matrices corresponding to the $\myShift{\delta}$ 
(i.e. $\vec\delta$-translations), 
\begin{equation}
  \myShift{\delta} \langle a | 
  = 
  \langle a | {\overline{\mymatrix{R}}\,}^{-1}, 
  \qquad
  \myShift{\delta} \langle b | 
  =  
  \langle b | \overline{\mymatrix{L}} 
\end{equation}
and are given by 
\begin{equation}
  \begin{array}{l}
  \overline{\mymatrix{R}}  
  = 
  \left[ 
    \prod_{\ell=0}^{3} \left( \mymatrix{R} - \alpha_{\ell} \right) 
  \right]^{1/4}, 
  \\[1mm]
  \overline{\mymatrix{L}}  
  = 
  \left[ 
    \prod_{\ell=0}^{3} \left( \mymatrix{L} - \alpha_{\ell} \right) 
  \right]^{1/4}. 
  \end{array} 
\end{equation}
These matrices and rows satisfy 
\begin{equation}
  \mymatrix{L}^{\pm} \mymatrix{A}^{\pm} 
  - 
  \mymatrix{A}^{\pm} \mymatrix{L}^{\mp} 
  = 
  |1\rangle \langle a^{\pm} | 
\end{equation}
and their dependence on $\vec{n}$ can be presented as 
\begin{equation}
  \begin{array}{l}
  \langle a^{\pm}(\vec{n}) | 
  = 
  \langle c^{\pm} | \mymatrix{X}^{\pm}(\vec{n}), 
  \\[1mm]
  \mymatrix{A}^{\pm}(\vec{n}) 
  =
  \mymatrix{C}^{\pm} \mymatrix{X}^{\pm}(\vec{n}) 
  \end{array}
\end{equation}
where 
\begin{equation}
  \mymatrix{X}^{\pm}(\vec{n}) 
  = 
  \prod_{i=1}^{3} \left( \mymatrix{X}^{\pm}_{i}\right)^{n_{i}} 
\label{bright:def-X}
\end{equation}
with 
\begin{equation}
  \mymatrix{X}^{\pm}_{i} 
  = 
  \left( \mymatrix{L}^{\mp} - \alpha_{i} \right)^{\pm 1} 
  \left( \mymatrix{L}^{\mp}_{*} \right)^{\mp 1} 
\label{bright:def-Xi}
\end{equation}
and 
\begin{equation}
  \mymatrix{L}^{\pm}_{*}  
  = 
  \left[ 
    \left( \mymatrix{L}^{\pm} - \alpha_{0} \right)^{-1} 
    \prod\nolimits_{i=1}^{3} \left( \mymatrix{L}^{\pm} - \alpha_{i} \right) 
  \right]^{1/2}. 
\end{equation}

Again, as in the `dark'-soliton case, the appearance of the fractional powers 
in the above formulae leads to some restrictions on the parameters 
$L_{n}^{\pm}$ ($n=1, ..., N$): 
\begin{equation}
  \prod\nolimits_{\ell=0}^{3} 
  \left( L_{n}^{\pm} - \alpha_{\ell} \right) 
  > 0 
  \qquad
  ( n = 1, ..., N). 
\label{bright:restr}
\end{equation}
The analysis of these inequalities is easier then that of the corresponding 
ones, \eref{dark:restr}, from the previous section. 
The simplest (but not the only) 
solution is to calculate $\alpha_{0}$ from \eref{restr:lambda} and then take 
all $L_{n}^{\pm}$ greater then 
$\max\limits_{\ell=0, ..., 3} \alpha_{l}$. 

Finally, we can formulate the main result of this section as follows.

\begin{propositionA} \label{prop:bright}
The `bright'-soliton solutions can be presented as 

\begin{equation}
  u(\vec{n}) 
  = 
  \langle a^{\pm}(\vec{n}) | 
  \left[ 
    \mymatrix{1} 
    + 
    \mymatrix{U}^{\pm}(\vec{n}) 
  \right]^{-1} 
  | \,1\, \rangle
\qquad
  (\myin{\pm})
\end{equation}
where 
\begin{equation}
  \mymatrix{U}^{\pm}(\vec{n}) 
  = 
  \mymatrix{A}^{\mp}(\vec{n}) \, 
  \mymatrix{M}^{\pm} \, 
  \mymatrix{A}^{\pm}(\vec{n}) 
\end{equation}
with 
\begin{equation}
  \mymatrix{M}^{\pm} 
  = 
  \left[ 
    \prod\nolimits_{\ell=0}^{3} 
    \left( \mymatrix{L}^{\pm} - \alpha_{\ell} \right) 
  \right]^{1/2} 
\end{equation}
and 
\begin{equation}
  \langle a^{\pm}(\vec{n}) | 
  = 
  \langle c^{\pm} | \mymatrix{X}^{\pm}(\vec{n}), 
\end{equation}
\begin{equation}
  \mymatrix{A}^{\pm}(\vec{n}) 
  =
  \mymatrix{C}^{\pm} \mymatrix{X}^{\pm}(\vec{n}) 
\end{equation}
where $\langle c^{\pm} |$ are constant $N$-rows, 
$\langle c^{\pm} | = ( c^{\pm}_{1}, ..., c^{\pm}_{N} )$ 
and $\mymatrix{C}^{\pm}$ are constant matrices given by 
\begin{equation}
  \mymatrix{C}^{\pm} 
  = 
  \left(\; 
    \frac{ c^{\pm}_{n} }{ L^{\pm}_{m} - L^{\mp}_{n} }
  \;\right)_{m,n=1, ..., N}   
\end{equation}
with the matrices $\mymatrix{X}^{\pm}(\vec{n})$ being defined in 
\eref{bright:def-X} and \eref{bright:def-Xi}.

Here, 
$\alpha_{0} = \alpha_{0}\left(\alpha_{1},\alpha_{2},\alpha_{3}\right)$  
is given by 
\begin{equation}
  \alpha_{0} 
  = 
  \sum_{i=1}^{3} \alpha_{i} 
  - 
  \frac{ \sum_{i=1}^{3} \Gamma_{i} \alpha_{i}^{2}  } 
       { \sum_{i=1}^{3} \Gamma_{i} \alpha_{i} } 
\end{equation}
and 
$c_{n}^{\pm}$, $L_{n}^{\pm}$ ($n=1, ..., N$) and 
$\alpha_{i}$ ($i=1,2,3$) are arbitrary (up to the restrictions 
\eref{bright:restr}) constants.

\end{propositionA}

The simplest, 1-soliton solution can be presented as 
\begin{equation}
  \myu(\vec{n}) 
  =
  \pm \, u_{*} \, 
  \frac{ e^{ \pm f(\vec{n}) } } 
       { \cosh\left( h(\vec{n}) \pm \delta \right) } 
\qquad
  (\myin{\pm})
\end{equation}
where 
\begin{equation}
  u_{*} 
  = 
  \frac{ \left| L^{+} - L^{-} \right| } 
       { 2 \sqrt{ \bar{L}^{+} \bar{L}^{-} } }, 
\end{equation}
\begin{equation}
  f(\vec{n}) = f_{0} + \left( \vec{\varphi}, \vec{n} \right), 
  \qquad 
  h(\vec{n}) = h_{0} + \left( \vec{\chi}, \vec{n} \right) 
\end{equation}
with arbitrary $f_{0}$ and $h_{0}$ and  
\begin{equation}
  \begin{array}{lcl}
  \vec{\varphi}  
  =  
  \sum_{i=1}^{3} \varphi_{i} \vec{e}_{i}, 
  &&
  \varphi_{i} 
  = 
  \frac{1}{2} 
  \ln \, 
    \frac{ L^{-} - \alpha_{i} }{ L^{-}_{*} } \, 
    \frac{ L^{+} - \alpha_{i} }{ L^{+}_{*} } 
  \\[4mm]
  \vec{\chi}  
  =  
  \sum_{i=1}^{3} \chi_{i} \vec{e}_{i}, 
  &&
  \chi_{i} 
  = 
  \frac{1}{2} 
  \ln \, 
    \frac{ L^{-} - \alpha_{i} }{ L^{+} - \alpha_{i} } \, 
    \frac{ L^{+}_{*} }{ L^{-}_{*} }. 
  \end{array}
\end{equation}
Here, the two constants $L^{\pm}$ replace the matrices $\mymatrix{L}^{\pm}$, 
the constant $\delta$ which is used instead of the matrices 
$\mymatrix{M}^{\pm}$ is given by 
$ \delta = \frac{1}{2} \ln \frac {\bar{L}^{+}}{\bar{L}^{-}} $ 
with the `averages' $\bar{L}^{\pm}$ and $L^{\pm}_{*}$ defined as 
\begin{equation}
  \bar{L}^{\pm}  
  = 
  \left[ 
    \prod\nolimits_{\ell=0}^{3} \left( L^{\pm} - \alpha_{\ell} \right) 
  \right]^{1/4}, 
\qquad 
  L^{\pm}_{*}  
  = 
  \left[ 
    \frac{ \prod_{i=1}^{3} \left( \mymatrix{L}^{\pm} - \alpha_{i} \right) }
         { L^{\pm} - \alpha_{0} } 
  \right]^{1/2}. 
\end{equation}
Again, as in the `dark'-soliton case, we have a real analogue of the bright 
soliton of the complex models: the plane wave 
$\pm \, u_{*} \, e^{ \pm f(\vec{n}) }$ 
modulated by the typical soliton factor (this time the sech-factor) 
which vanishes in the $\left| \vec{n} \right| \to \infty$ limit, 
except in the cases when we approach the infinity along the directions 
perpendicular to $\vec{\chi}$. 
  
To conclude this section, we would like to note that using equation 
\eref{quasi-ansatz}, without introducing the $\vec{\alpha}$-vectors, one arrives 
at the restrictions
$ 
  \prod_{\ell=0}^{3} 
  \left( \mymatrix{L} - g_{\ell} \right)
  = 
  \prod_{\ell=0}^{3} 
  \left( \mymatrix{R} - g_{\ell} \right)
  = 
  \mymatrix{1} 
$ 
where $\{ g_{\ell} \}_{\ell=0}^{3}$ are parameters corresponding to 
$\{ \vec{g}_{\ell} \}_{\ell=0}^{3}$. Thus, for a given set of the 
$g$-parameters, one has to construct two diagonal matrices of only four roots 
of the fourth-order  equation, which leads to general 2-soliton solutions or 
degenerate solutions with one soliton in one component 
(say, $\mymatrix{L} \propto \mymatrix{1}$) and three solitons in another one, 
whereas proposition \ref{prop:bright}, resulting from \eref{def:alpha} gives 
$N$-soliton solutions for arbitrary $N$.

\section{Conclusion.} 

As one can see from the above presentation, the procedure of deriving the 
$N$-soliton solutions was mostly the reduction to the already known equations: 
we have established the links between the field equations \eref{eq:main} and 
the Hirota-Ablowitz-Ladik system \eref{syst:hal}. 

We would like to note once more that there were two non-trivial steps in our 
algorithm. First was the two-sublattice representation of $\myu(\vec{n})$ 
given by \eref{def:qr} which has been used in our previous works 
\cite{V16a,V16b} and which may be viewed as the most straightforward way to 
introduce the Ablowitz-Ladik triplet of the tau-functions. 

The second moment was the introduction of the frame 
$\{ \vec{\alpha}_{\ell} \}_{\ell=0}^{3}$. In this paper we have used it just 
to split the field equations and, actually, as a way to resolve the 
restriction imposed on translations (for example, that translations 
corresponding to $\vec{e}_{i}$ and $-\vec{e}_{i}$ are mutually inverse). 
However, this construction, whose geometric importance has been discussed by 
various authors (see, for example, \cite{ABS12}), can be generalized to connect 
other Hirota-type (star-type) equations that are typical for physical 
applications with the cell-type (defined, for example, on a cube or an 
octahedron) equations that usually appear in the mathematical works devoted to 
such questions as classification, integrability, geometric content etc.

Considering the restriction \eref{restr:gamma} we would like to note that it 
was crucial for the procedure we have used to \emph{derive} the presented 
solutions. However, we cannot claim that it is necessary for the 
existence of the soliton-like solutions or the integrability of the model. 
Restrictions of this type often appear in the studies of integrable models. 
If we consider, for example, the HBDE, the restriction similar to 
\eref{restr:gamma} is present in the most  of the works devoted to this system 
(including the original paper \cite{H81}). However, as it has been demonstrated 
in, for example, \cite{RGS92}, the HBDE is integrable even without it 
(the widespread opinion now is that it is required for the existence of 
Hirota-form soliton solutions).
At the same time, there are many known situations when the integrability, and 
hence the existence of the solitons, are related to some restrictions on the 
constants of a model. So, the role of the restriction \eref{restr:gamma} 
remains an open question.

The fact that model \eref{def:model} possesses the N-soliton solutions is a 
strong evidence (but surely not a proof) of its integrability. Thus, a 
straightforward continuation of this work is to look for the zero-curvature 
representation, the conservation laws, the B\"acklund transformations etc, 
i.e. to analyze the standard set of problems that arises in connection with 
any integrable system. 
However, these questions are out of the scope of the present paper and surely 
deserve separate studies.

\ack 

We would like to thank the referee for the constructive comments 
and suggestions for improvement of this paper.


\appendix

\section{Proof of \eref{restr:lambda}. \label{app:lambda}} 

By simple algebra, one can derive from \eref{def:hat-Gamma} the identity 
\begin{equation}
  \hat\Gamma_{i} \, 
  \prod_{ \stackrel{\ell=0 }{\scriptscriptstyle (\ell \ne i) } }^{3}
  ( \mylambda{i} - \mylambda{\ell} ) \; 
  = 
    C_{i} \, G_{0} 
  + \left( \mylambda{0} - L \right) G_{1} 
  + G_{2} 
\label{appa:hatg}
\end{equation}
where 
\begin{equation}
  L = \sum_{i=1}^{3} \mylambda{i}, 
  \qquad 
  G_{n} = \sum_{i=1}^{3} \Gamma_{i} \mylambda[n]{i} 
  \quad (n=0,1,2) 
\end{equation}
and 
\begin{equation}
  C_{i} 
  = 
  \mylambda[2]{i} 
  - 
  \mylambda{0} 
  \mylambda{i} 
  + 
  \mylambda{j} 
  \mylambda{k} 
  \qquad \myijk. 
\end{equation}

In our case, due to the restriction \eref{restr:gamma}, $G_{0}=0$. 
Thus, the right-hand side of \eref{appa:hatg} (and hence all $\hat\Gamma_{i}$) 
vanishes when 
\begin{equation}
  \mylambda{0} 
  = L - G_{2} / G_{1} 
\end{equation}
which proves \eref{restr:lambda}.

\section{Proof of proposition \ref{dark:prop-hal}. \label{app:dark}} 

To prove the fact that functions $\tau$ and $\rho$ 
defined in proposition \ref{dark:prop-hal} 
satisfy \eref{syst:alh-r} 
one has just to rewrite \eref{dark:fay} with $\zeta=0$ 
\begin{equation}
  0 = 
  (\xi - \eta) \myShifted[1]{\xi\eta}{\Omega} \myShifted[1]{0}{\Omega}
  - \xi \myShifted[1]{\xi 0}{\Omega} \myShifted[1]{\eta}{\Omega}
  + \eta \myShifted[1]{\eta 0}{\Omega} \myShifted[1]{\xi}{\Omega}
\label{appb:fay}
\end{equation}
and to express $\Omega$ and $\myShifted{0}{\Omega}$ in terms of 
$\tau$ and $\rho$.

Applying $\myShift{0}^{-1}$ to \eref{appb:fay} 
and expressing $\Omega$ and $\myShifted[2]{0}{\Omega}$ in terms of 
$\tau$ and $\sigma$ one 
can see that $\tau$ and $\sigma$ satisfy \eref{syst:alh-s} 
with $\myalhconst{a}{\xi,\eta} = \xi-\eta$.

Finally, to prove \eref{syst:hal-t}, we rewrite 
\eref{appb:fay} with $\eta$ replaced with $1/\eta$:
\begin{equation}
\fl\qquad
  0 = 
  (\xi\eta-1) 
  \myShifted[1]{\xi}{\myShifted{1/\eta}{\Omega}} 
  \myShifted[1]{0}{\Omega} 
  - 
  \xi\eta 
  \myShifted[1]{\xi 0}{\Omega} 
  \myShifted[1]{1/\eta}{\Omega}
  + 
  \myShifted[1]{1/\eta}{\myShifted{0}{\Omega}} 
  \myShifted[1]{\xi}{\Omega}. 
\end{equation}
After application of $\myShift{1/\eta}^{-1}$, which is equal, due to 
\eref{dark:duality}, to $\myShift{\eta} \myShift{0}^{-1}$, this identity 
becomes
\begin{equation}
  0 = 
  (\xi\eta-1) 
  \myShifted[1]{\xi}{\Omega} 
  \myShifted[1]{\eta}{\Omega} 
  - 
  \xi\eta 
  \myShifted[1]{\xi\eta}{\Omega} 
  \Omega 
  + 
  \myShifted[1]{0}{\Omega} 
  \myShifted[1]{\xi\eta}{\myShifted[2]{0}{\Omega}}. 
\end{equation}
Replacing $\Omega$ and $\myShift{0}^{\pm 1} \Omega$ with 
$\tau$, $\rho$ and $\sigma$ one arrives at \eref{syst:alh-t} with 
$\myalhconst{b}{\xi,\eta} = 1 - 1/\xi\eta$. 

This concludes the proof of proposition \ref{dark:prop-hal}.


\end{document}